\pdfoutput=1
\documentclass[sigplan,screen]{acmart}
\setcopyright{none}

\acmConference{Preprint}{Jan. 2021}{Online}
\acmYear{2021}
\acmISBN{} 
\acmDOI{} 
\startPage{1}

\usepackage{booktabs}%
\usepackage{subcaption}%
\usepackage[utf8]{inputenc}
\usepackage[T1]{fontenc}
\usepackage[english]{babel}
\usepackage{fancyvrb}
\usepackage{graphicx}
\usepackage{csquotes}
\usepackage{microtype}
\usepackage{xspace}
\usepackage{syntax}
\usepackage{mathpartir}

\usepackage{tikz}
\usetikzlibrary{calc}
\usetikzlibrary{shadows,graphs}
\tikzstyle{Box}=[draw, fill=gray!20, text centered]
\usepackage{pgfplots}

\usepackage{amsthm}
\usepackage{MnSymbol}
\theoremstyle{definition}

\newtheorem*{thm}{Theorem}
\theoremstyle{remark}

\newcommand{\mlang}{\textsc{Mlang}\xspace}
\newcommand{\mpp}{\mbox{M\hspace{-.05em}\raisebox{.2ex}{+}\raisebox{.2ex}{+}}\xspace}
\newcommand{\ocaml}{OCaml\xspace}
\let\dgfip\DGFiP
\newcommand{\sref}[1]{Section~\ref{sec:#1}}
\newcommand{\fref}[1]{Fig.~\ref{fig:#1}}

\newcommand{\myparagraph}[1]{\textbf{\textit{#1.}}}
\newcommand{\mypointlessparagraph}[1]{\textbf{\textit{#1}}}

\usepackage{listings}
  
\usepackage{xparse}
  \NewDocumentCommand{\li}{v}{\lstinline{#1}}

\begin{document}

\title{A Modern Compiler for the French Tax Code}%
\author{Denis Merigoux}
\authornote{Equal contribution}
\orcid{0000−0003−2247−0938}%
\affiliation{
	\institution{Inria}%
	\city{Paris}
	\country{France}%
}
\email{denis.merigoux@inria.fr}%

\author{Raphaël Monat}
\authornotemark[1]
\orcid{0000-0001-8487-0326}             
\affiliation{
  \institution{Sorbonne Université, CNRS, LIP6}
  \streetaddress{F-75005}
  \city{Paris}
  \country{France}%
}
\email{raphael.monat@lip6.fr}%

\author{Jonathan Protzenko}
\affiliation{
	\institution{Microsoft Research}%
	\country{USA}%
}
\email{protz@microsoft.com}%

\begin{abstract}

  In France, income tax is computed from taxpayers' individual returns, using an
  algorithm that is authored, designed and maintained by the French Public
  Finances Directorate (\dgfip).
  This algorithm relies on a legacy custom language and compiler originally
  designed in 1990, which unlike French wine, did not age well with time.
  Owing to the shortcomings of the input language and the technical limitations
  of the compiler, the algorithm is proving harder and harder to maintain,
  relying on ad-hoc behaviors and workarounds to implement the most recent
  changes in tax law. Competence loss and aging code also mean that the system
  does not benefit from any modern compiler techniques that would increase
  confidence in the implementation.

  We overhaul this infrastructure and present \mlang, an open-source compiler
  toolchain whose goal is to replace the existing infrastructure. \mlang
  is based on a reverse-engineered formalization of the \dgfip's system, and has
  been thoroughly validated against the private \dgfip test suite. As such,
  \mlang has a formal semantics; eliminates previous hand-written workarounds in
  C; compiles to modern languages (Python); and enables a variety
  of instrumentations, providing deep insights about the essence of French income
  tax computation. The \dgfip is now officially transitioning
  to \mlang for their production system.
\end{abstract}



\maketitle

\section{Introduction}
\label{sec:intro}
The French Tax Code is a body of legislation amounting to roughly 3,500 pages of text,
defining the modalities of tax collection by the state. In particular, each new
fiscal year, a new edition of the Tax Code describes in natural language
how to compute the final amount of income tax (IR, for \emph{impôt sur le
revenu}) owed by each household.

As in many other tax systems around the world, this computation is quite
complex. France uses a bracket system (as in, say, the US federal income tax),
along with a myriad of tax credits, deductions, optional rules, state-sponsored
direct aid, all of which are parameterized over the composition of the
household, that is, the number of children, their respective ages, potential
disabilities, and so on.

Unlike, say, the United States, the French system relies heavily on automation.
During tax season, French taxpayers log in to the online tax portal, which is
managed by the state. There, taxpayers are presented with online forms,
generally pre-filled. If applicable, taxpayers can adjust the forms, e.g. by
entering extra deductions or credits. Once the taxpayer is satisfied with the
contents of the online form, they send in their return. Behind the scenes, the
IR algorithm is run, and taking as input the contents of the forms, returns the
final amount of tax owed. The taxpayer is then presented
with the result at tax-collection time.

Naturally, the ability to independently reproduce and thus trust the IR
computation performed by the \dgfip is crucial. First, taxpayers need to
\emph{understand} the result, as their own estimate may differ (explainability).
Second, citizens may want to \emph{audit} the algorithm, to ensure it faithfully
implements the law (correctness). Third, a standalone, reusable implementation
allows for a complete and precise \emph{simulation} of the impacts of a tax reform,
greatly improving existing efforts~\cite{landais2011,leximpact} (forecasting).

Unfortunately, we are currently far from a transparent, open-source,
reproducible computation. Following numerous requests (using a disposition
similar to the United States' Freedom of Information Act), parts of the existing
source code were published. In doing so, the public learned that \emph{i)} the existing
infrastructure is made up of various parts pieced together and that \emph{ii)} key data
required to accurately reproduce IR computations was not shared with the public.

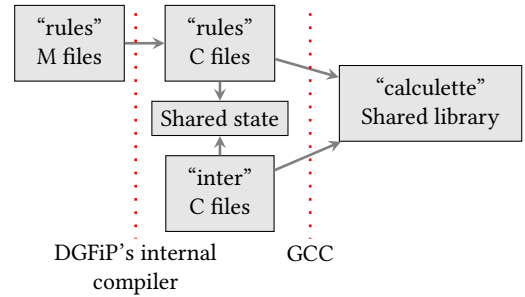
\begin{figure}
  \begin{center}
    \begin{tikzpicture}[scale=0.8,font=\smaller]
      \node (mrules) [Box] {$\begin{array}{c}\text{\enquote{rules}}\\\text{M files}\end{array}$};
      \node (crules) [Box] at ($ (mrules)+(2.5,0) $) {$\begin{array}{c}\text{\enquote{rules}}\\\text{C files}\end{array}$};
      \node (inter) [Box] at ($ (crules)+(0,-2.5) $) {$\begin{array}{c}\text{\enquote{inter}}\\\text{C files}\end{array}$};
      \node (shared) [Box] at ($ (crules)+(0,-1.25) $) {\text{Shared state}};
      \node (dll) [Box] at ($ (crules)+(3.5,-1) $) {$\begin{array}{c}\text{\enquote{calculette}}\\\text{Shared library}\end{array}$};

      \draw[-stealth,color=gray,line width=1pt] (mrules) -- (crules);
      \draw[-stealth,color=gray,line width=1pt] (crules) -- (dll);
      \draw[-stealth,color=gray,line width=1pt] (inter) -- (dll);

      \draw[-stealth,color=gray,line width=1pt] (crules) -- (shared);
      \draw[-stealth,color=gray,line width=1pt] (inter) -- (shared);

      \draw[loosely dotted, color=red, line width=1] ($ (mrules) + (1.1, 0.5) $) -- ($ (mrules)+(1.1,-3.25) $);
      \draw[loosely dotted, color=red, line width=1] ($ (crules) + (1.5, 0.5) $) -- ($ (crules)+(1.5,-3.25) $);

      \node (compir) at ($ (mrules)+(1.1,-3.75) $) { $\begin{array}{c}\text{DGFiP's internal}\\\text{compiler}\end{array}$ };
      \node (gcc) at ($ (crules)+(1.5,-3.5) $) { GCC };

      \end{tikzpicture}
  \end{center}
  \caption{Legacy architecture}
  \label{fig:dgfip:architecture}
\end{figure}

The current, legacy architecture of the IR tax system is presented in
\fref{dgfip:architecture}. The bulk of the tax code is described as a set of \enquote{rules}
authored in M, a custom, non Turing-complete language. A total of 90,000 lines
of M rules compile to 535,000 lines of C (including whitespace and comments) via
a custom compiler. Rules are now mostly public \cite{ir-published}.
Over time, the expressive power of rules turned out to be too
limited to express a particular feature, known as \emph{liquidations multiples},
which involves tax returns across different years. Lacking the
expertise to extend the M language, the \dgfip added in 1995 some high-level glue
code in C, known as \enquote{inter}. The glue code is closer to a full-fledged
language, and has a non-recursive call-graph which may call the
\enquote{rules} computation multiple times with various parameters. The \enquote{inter}
driver amounts to 35,000 lines of C code and has not been released.

Both \enquote{inter} and \enquote{rules} are updated every year to follow updates in the
law, and as such, have been extensively modified over their 30-year operational
lifespan.

Our goal is to address these shortcomings by bringing the French tax code
infrastructure into the 21\textsuperscript{st} century. Specifically, we wish to: \emph{i)}
reverse-engineer the unpublished parts of the \dgfip computation, so as to \emph{ii)}
provide an explainable, open-source, \emph{correct} implementation that can be
independently audited; furthermore, we wish to \emph{iii)} modernize the compiler
infrastructure, eliminating in the process any hand-written C that could not be
released because of security concerns, thus enabling a host of modern
applications, simulations and use-cases.

\begin{itemize}
  \item We start with a reverse-engineered formal semantics for the M DSL, along
    with a proof of type safety performed using the
    Coq~\cite{coq} proof assistant (\sref{semantics}).
  \item To eliminate C code from the ecosystem, we extend the M language with
    enough capabilities to encode the logic of the high-level ``inter'' driver
    (\fref{dgfip:architecture}) -- we dub the new design \mpp (\sref{mpp}).
  \item To execute M/\mpp programs, we introduce \mlang, a complete
    re-implementation which combines a
    reference interpreter along with an optimizing compiler
    that generates C and Python code (\sref{exec}).
  \item We evaluate our implementation: we show how we attained 100\%
    conformance on the legacy system's testsuite, then proceed to enable a variety of
    analyses and instrumentations to fuzz, measure and stress-test our new
    system (\sref{eval}).
  \item We conclude with a \emph{tour d'horizon} of related attempts
    at increasing trust in algorithmic parts of the law (\sref{conclusion}).
\end{itemize}

Our code is open-source and available on GitHub \cite{mlang-github} and as an archived artifact on Zenodo \cite{mlang-zenodo}.
We have engaged with the \dgfip, and following numerous
discussions, iterations, and visits to their offices, we have been formally
approved to start replacing the legacy infrastructure with our new
implementation, meaning that within a few years' time, all French tax returns
will be processed using the compiler described in the present paper.

\section{Giving Semantics to the M Language}
\label{sec:semantics}

\newcommand{\indef}{\lit*{undef}\xspace}
\newcommand{\farr}{\texttt{round}}
\newcommand{\finf}{\texttt{truncate}}
\newcommand{\fabs}{\texttt{abs}}
\newcommand{\fpos}{\texttt{pos}}
\newcommand{\fposo}{\texttt{pos\_or\_null}}
\newcommand{\fnull}{\texttt{null}}
\newcommand{\fpresent}{\texttt{present}}
\newcommand{\fmin}{\texttt{min}}
\newcommand{\fmax}{\texttt{max}}
\newcommand{\fplus}{\texttt{+}}
\newcommand{\fminus}{\texttt{-}}
\newcommand{\ftimes}{\texttt{*}}
\newcommand{\fdiv}{\texttt{/}}
\newcommand{\fand}{\texttt{\&\&}}
\renewcommand{\for}{\texttt{||}}
\newcommand{\fleq}{\texttt{<=}}
\newcommand{\feq}{\texttt{==}}
\newcommand{\mum}{$\mu$M\xspace}
\newcommand{\V}{\mathcal{V}}
\newcommand{\F}{\mathbb{F}}
\newcommand{\B}{\mathbb{B}}
\newcommand{\Ne}{\mathbb{N}}
\newcommand{\tabletype}[2]{[\![#1\;;\;#2]\!] }
\newcommand{\floatlbl}{float}
\newcommand{\programlbl}{program}
\newcommand{\cmdlbl}{command}
\newcommand{\varlbl}{var}
\newcommand{\exprlbl}{expr}
\newcommand{\compoplbl}{compop}
\newcommand{\compop}{\synt{\compoplbl}}
\newcommand{\binoplbl}{binop}
\newcommand{\unoplbl}{unop}
\newcommand{\unop}{\synt{unop}}
\newcommand{\arithoplbl}{arithop}
\newcommand{\booloplbl}{boolop}
\newcommand{\errorlbl}{error}
\newcommand{\minuslbl}{minus}
\newcommand{\notlbl}{not}
\newcommand{\condlbl}{conditional}
\newcommand{\iflbl}{if}
\newcommand{\thenlbl}{then}
\newcommand{\elselbl}{else}
\newcommand{\eerror}{\synt{\errorlbl}}
\newcommand{\eif}{\lit*{\iflbl}}
\newcommand{\ethen}{\lit*{\thenlbl}}
\newcommand{\eelse}{\lit*{\elselbl}}
\newcommand{\arithop}{\synt{\arithoplbl}}
\newcommand{\boolop}{\synt{\booloplbl}}
\newcommand{\binop}{\synt{\binoplbl}}
\newcommand{\float}{\synt{\floatlbl}}
\newcommand{\valeur}{\synt{value}}
\newcommand{\treal}{\lit*{\reallbl}}
\newcommand{\undefi}{\lit*{indéfini}}
\newcommand{\tvaleur}{\lit*{scalar}}
\newcommand{\ttableau}{\lit*{array}}
\newcommand{\typed}[2]{#1 : #2}
\newcommand{\ftyped}[3]{#1 : #2 \rightarrow #3}
\newcommand{\program}{\synt{\programlbl}}
\newcommand{\jprogram}[3]{#1 \vdash #2 \Rrightarrow #3}
\newcommand{\cmd}{\synt{\cmdlbl}}
\newcommand{\jcmd}[3]{#1 \vdash #2 \Rrightarrow #3}
\newcommand{\expr}{\synt{\exprlbl}}
\newcommand{\texpr}[2]{#1\vdash #2\;}
\newcommand{\trule}[1]{T-#1}
\newcommand{\drule}[1]{D-#1}
\newcommand{\valued}[2]{#1\Downarrow #2}
\newcommand{\dexpr}[4]{#2\vdash\valued{#3}{#4}}%
\newcommand{\dcmd}[4]{#2\vdash#3\Rrightarrow #4}
\newcommand{\dprog}[4]{#2\vdash#3\Rrightarrow #4}

The 2018 version of the income tax computation \cite{ir-published} is split across 48 files, for a total of 92,000 lines of code. The code is
written in M,
the input language originally designed by the \dgfip. In order to understand
this body of tax code, we set out to give a semantics to M.

\subsection{Overview of M}
\label{sec:overview}

M programs are made up of two parts: declarations and rules.

\emph{Declarations} introduce: input variables, intermediary variables, output variables
and exceptions. Variables are either scalars or fixed-length arrays. Both
variables and exceptions
are annotated with a human-readable description.
Variables that belong to the same section of the tax form are annotated with the
same kind. Examples of kinds include "triggers tax credit", or "is advance payment".
This is used later in \mpp (\sref{mpp:example}) for partitioning variables,
and quickly checking whether any variable of a given kind has a non-\indef
value.

\emph{Rules} capture the computational part of an M program; they are either
variable assignments or raise-if statements.

As a first simplified example, the French tax code declares an input variable
\li+V_0AC+ for whether an individual is single (value \li+1+) or not (value \li+0+).
Lacking any notion of data type or enumeration, there is no way to enforce
statically that an individual cannot be married (\li+V_0AM+) and single
(\li+V_0AC)+ at the same time.
Instead, an exception \li+A031+ is declared, along with a
human-readable description. Then, a rule raises an exception if the sum of the
two variables is greater than 1. (The seemingly superfluous
\li!+ 0! is explained in \sref{mum:sem}.)
For the sake of example, we drop irrelevant extra syntactic features, and for the sake
of readability, we translate keywords and descriptions into English.

{\footnotesize
\begin{verbatim}
V_0AC : input family ... : "Checkbox : Single" type BOOLEAN ;
V_0AM : input family ... : "Checkbox : Married" type BOOLEAN ;
A031:exception :"A":"031":"00":"both married and single":"N";

if V_0AC + V_0AM + 0 > 1 then error A031 ;
\end{verbatim}
}

As a second simplified example, the following M rule computes the value of a
hypothetical variable
\li+TAXBREAK+. Its value is computed from
variables \li+CHILDRENCOUNT+ (for the number of children in the household) and
\li+TAXOWED+ (for the tax owed before the break) -- the assigned expression relies on a conditional and
the built-in \li+max+ function. This expression gives a better tax break
to households having three or more children.

{\footnotesize
\begin{verbatim}
TAXBREAK= if (CHILDRENCOUNT+0 > 2)
  then max(MINTAXBREAK,TAXOWED * 20 / 100)
  else MINTAXBREAK endif;
\end{verbatim}
}

\noindent
For the rest of this paper, we abandon concrete syntax and all-caps
variable names, in favor of a core calculus that faithfully models M: \mum.

\subsection{\mum: a core model of M}

The \mum core language omits variable
declarations, whose main purpose is to provide a human-readable description
string that relates them to the original tax form.
The \mum core language also eliminates syntactic sugar, such as statically
bounded loops, or type aliases (e.g. \li+BOOLEAN+).
Finally, a particular feature of M is that rules may be provided in any order:
the M language has a built-in dependency resolution feature that automatically
re-orders computations (rules) and asserts that there are no loops in variable
assignments. In our own implementation (\mlang, \sref{exec}), we perform a topological sort;
in our \mum formalization, we assume that computations are already in a suitable
order.

\subsection{Syntax of \mum}

We describe the syntax of \mum in \fref{mum:syntax}. A program is a
series of statements (\enquote{rules}).  Statements are either raise-error-if, or assignments.
We define two forms of assignment: one
for scalars and the other for fixed-size arrays. The latter is of the form
\texttt{a[X, n] := e}, where \lit*{X} is bound in \texttt{e}
(the index is \emph{always} named \lit*{X}). Using Haskell's list comprehension
syntax, this is to be understood as $a := [ e | X \leftarrow [0 .. n - 1] ]$.

Expressions are a combination of
variables (including the special index expression \lit*{X}), values,
comparisons, logic and arithmetic expressions, conditionals, calls to builtin
functions, or index accesses. Most functions exhibit standard behavior on
floating-point values, but M assumes the default IEEE-754 rounding mode, that
is, rounding to nearest and ties to even.
The detailed behavior of each function is described in
\fref{mum:sem:func}.

Values can be \indef, which arises in two situations: references to
variables that have not been defined (i.e.
for which the entry in the tax form has been left blank) and out of bounds
array accesses.
All other values are IEEE-754 double-precision numbers, i.e.
64-bit floats. The earlier \li+BOOLEAN+ type (\sref{overview}) is
simply an alias for a float whose value is implicitly \li+0+ or \li+1+. There is no other
kind of value, as a reference to an array variable is invalid.
Function {\fpresent} discriminates the \indef value from floats.

\begin{figure}%
  {\footnotesize
	\begin{grammar}
		<\programlbl> ::= <\cmdlbl> | <\cmdlbl> ";" <\programlbl>

		<\cmdlbl> ::= "\iflbl" <\exprlbl> "\thenlbl" <error>
                \alt <\varlbl> ":=" <\exprlbl> | <\varlbl> "[" "X" ";" <\floatlbl> "]" ":=" <\exprlbl>

		<\exprlbl> ::= <\varlbl> | \lit*{X} | <value> |  <\exprlbl> <\binoplbl>  <\exprlbl>
                \alt <\unoplbl> <\exprlbl> |"\iflbl" <\exprlbl> "\thenlbl" <\exprlbl> "\elselbl" <\exprlbl>
		\alt <func> "(" <\exprlbl>, \ldots, <\exprlbl> ")" | <\varlbl> "[" <\exprlbl> "]"

		<value> ::= "\indef" | <\floatlbl>

		<\binoplbl> ::= <\arithoplbl>  | <\booloplbl>

		<\arithoplbl> ::= "+" | "-" | "*" | "/"

		<\booloplbl> ::= "<=" | "<" | ">" | ">=" | "==" | "!=" | "&&" | "||"

		<\unoplbl> ::= "-" | "~"

		<func> ::= "\farr" | "\finf" | "\fmax" | "\fmin" | "\fabs"
		\alt "\fpos" | "\fposo" | "\fnull" | "\fpresent"
    \end{grammar}
    }
	\caption{Syntax of the \mum language}
    \label{fig:mum:syntax}
\end{figure}

\subsection{Typing \mum}

Types in \mum are either scalar or array types. M does not offer
nested arrays. Therefore, typing is mostly about making sure scalars and arrays
are not mixed up.

In \fref{typing}, a first judgment $\boxed{\texpr{\Gamma}{e}}$ defines
expression well-formedness. It rules out references to arrays, hence enforcing
that expressions have type scalar and that no values of type array can be
produced. Furthermore, variables may have no assignment at all (if the
underlying entry in the tax form has been left blank) but may still be referred
in other rules. Rather than introduce spurious variable assignments with
\indef, we remain faithful to the very loose nature of the M language and
account for references to undefined variables.

Then, $\boxed{\jprogram{\Gamma}{\program}{\Gamma'}}$ enforces well-formedness for
a whole program while returning an extended environment $\Gamma'$. We take
advantage of the fact that scalar and array assignments have different syntactic
forms. M disallows assigning different types to the same variable; we rule this
out in \TirName{T-Assign-*}.
A complete \mum program is well-formed if $ \jprogram\emptyset{P}{\_}$.

\begin{figure}%
	{\footnotesize
	\begin{align*}
		 & \text{Global function environment $\Delta$:} \\
		 & \Delta(\farr) = \Delta(\finf) = \Delta(\fabs) = \Delta(\fpos)\\
                 & = \Delta(\fposo) = \Delta(\fnull) = \Delta(\fpresent) =1\\
		 & \Delta(\fmin) = \Delta(\fmax) = \Delta(\arithop) = \Delta(\boolop) = 2 	\end{align*}
	\vspace{-1em}
	\[
		\text{Judgment : }
		\boxed{\texpr{\Gamma}{e}}\quad\text{(\enquote{Under $\Gamma$, $e$ is well-formed})}
	\]
	{
	\begin{mathpar}
		\inferrule[\trule{float}]{\quad}{\texpr{\Gamma}{\float}}\and
		\inferrule[\trule{undef}]{~}{\texpr{\Gamma}{\indef}}\and
		\inferrule[\trule{var-undef}]{x \not \in \text{dom }\Gamma}{\texpr{\Gamma}{x}}\and
		\inferrule[\trule{var}]{\Gamma(x) = \tvaleur}{\texpr{\Gamma}{x}}\and
		\inferrule[\trule{index-undef}]{
			x \not \in \text{dom }\Gamma\\
			\texpr{\Gamma}{e}\\
		}{
			\texpr{\Gamma}{x[e]}
		}
		\and
		\inferrule[\trule{conditional}]{
			\texpr{\Gamma}{e_1}\\
			\texpr{\Gamma}{e_2}\\
			\texpr{\Gamma}{e_3}
		}{
			\texpr{\Gamma}{\eif\;e_1\;\ethen\;e_2\;\eelse\;e_3}
		}\and
		\inferrule[\trule{index}]{
			\Gamma(x) = \ttableau\\
			\texpr{\Gamma}{e}
		}{
			\texpr{\Gamma}{x[e]}
		}\and
		\inferrule[\trule{func}]{
                        \Delta(f) = n\\
			\texpr{\Gamma}{e_1}\\
                        \cdots\\
			\texpr{\Gamma}{e_n}
		}{
			\texpr{\Gamma}{f(e_1,\ldots,e_n)}
		}
	\end{mathpar}}
	\vspace{1em}
	\centering
	\[
		\text{Judgment : }\text{ }
		\boxed{\jcmd{\Gamma}{\cmd}{\Gamma'}}
		\text{ and }
              \]
              \[
		\boxed{\jprogram{\Gamma}{\program}{\Gamma'}}
		\quad\text{(\enquote{$P$ transforms $\Gamma$ to $\Gamma'$})}
	\]
	\begin{mathpar}
          \inferrule[\trule{cond}]{
            \texpr{\Gamma}{e}
          }{
            \jcmd{\Gamma}{\eif\;e\;\ethen\;\eerror\;}{\Gamma}
          }\and
          \inferrule[\trule{seq}]{
            \jcmd{\Gamma_0}{c}{\Gamma_1}\\
            \jprogram{\Gamma_1}{P}{\Gamma_2}
          }{
            \jprogram{\Gamma_0}{c\;\lit*{;}\;P}{\Gamma_2}
          }\and
          \inferrule[\trule{assign-scalar}]{
            x \in \Gamma \Rightarrow \Gamma(x) = \tvaleur\\
            \texpr{\Gamma}{e}
          }{
            \jcmd{\Gamma}{x \;\lit*{:=}\; e\;}{\Gamma[x \mapsto \tvaleur]}
          }\and
          \inferrule[\trule{assign-array}]{
            x \in \Gamma \Rightarrow \Gamma(x) = \ttableau\\
            \texpr{\Gamma[\lit*{X} \mapsto  \tvaleur]}{e}
          }{
            \jcmd{\Gamma}{x[\lit*{X}, n]\;\lit*{:=}\;e\;}{\Gamma[x \mapsto \ttableau]}
          }\and
        \end{mathpar}}
      \caption{Typing of expressions and programs}
      \label{fig:typing}
\end{figure}

\subsection{Operational semantics of \mum}
\label{sec:mum:sem}

At this stage, seeing that there are neither unbounded loops nor user-defined
(recursive) functions in the language, M is obviously \emph{not} Turing-complete.
The language semantics are nonetheless quite devious, owing to the \indef
value, which can be explicitly converted to a float via a \li!+ 0!, as seen in
earlier examples. We proceed to formalize them in Coq~\cite{coq}, using the Flocq library \cite{flocq}. This ensures
we correctly account for all cases related to the \indef value, and guides
the implementation of \mlang (\sref{mpp}).

\myparagraph{Expressions}
The semantics of expressions is defined in \fref{mum:sem:expr}.
The memory environment, written $\Omega$ is a function from variables to either
scalar values (usually denoted $v$), or arrays (written $(v_0, \ldots,
v_{n-1})$).
A value absent from the environment evaluates to \indef.

The special array index variable \lit*{X} is evaluated as a normal
variable. Conditionals reduce normally, except when the guard is \indef:
in that case, the whole conditional evaluates into \indef.  If an index
evaluates to \indef, the whole array access is \indef.  In the case of a
negative out-of-bounds index access the result is 0; in the case of a positive
out-of-bounds index access the result is \indef.  Otherwise, the index is
truncated into an integer, used to access $\Omega$.
The behavior of functions,
unary and binary operators is described in \fref{mum:sem:func}.

Figuring out these (unusual) semantics took over a year.
We initially worked in a black-box setting, using as an oracle for our semantics
the simplified online tax simulator offered by the \dgfip. After the initial set
of M rules was open-sourced, we simply manually crafted test cases and fed those
by hand to the online simulator to adjust our semantics.  This allowed us to
gain credibility and to have the \dgfip take us seriously. After that, we were
allowed to enter the \dgfip offices and browse the source of their M compiler, as
long as we did not exfiltrate any information. This final ``code browsing''
allowed us to understand the ``inter'' part of their compiler, a well as nail
down the custom operators from \fref{mlang:customrounding}.

\begin{figure}%
  {\footnotesize
		\centering
		\[
			\text{Judgment : }
			\boxed{\dexpr{\Gamma}{\Omega}{e}{v}}
			\quad\text{(\enquote{Under $\Omega$, $e$ evaluates to $v$})}
		\]
		\begin{mathpar}
			\inferrule[\drule{value}]{v \in \valeur}{\dexpr{\Gamma}{\Omega}{v}{v}}\and
			\inferrule[\drule{var-undef}]{x \not \in \text{dom}~ \Omega}{\dexpr{\Gamma}{\Omega}{x}{\indef}}\and
			\inferrule[\drule{var}]{\Omega(x) = v}{\dexpr{\Gamma}{\Omega}{x}{v}}\and
			\and
			\inferrule[\drule{cond-true}]{
				\dexpr{\Gamma}{\Omega}{e_1}{f}\\
				f\notin \{0, \indef\}\\
				\dexpr{\Gamma}{\Omega}{e_2}{v_2}
			}{
                                \dexpr{\Gamma}{\Omega}{\eif\;e_1\;\ethen\;e_2\;\eelse\;e_3}{v_2}
                        }\and
                        \inferrule[\drule{X}]{\Omega(\lit*{X}) = v}{\dexpr{\Gamma}{\Omega}{\lit*{X}}{v}}
                        \and
			\inferrule[\drule{cond-false}]{
				\dexpr{\Gamma}{\Omega}{e_1}{0}\\
				\dexpr{\Gamma}{\Omega}{e_3}{v_3}
			}{
				\dexpr{\Gamma}{\Omega}{\eif\;e_1\;\ethen\;e_2\;\eelse\;e_3}{v_3}
			}\and
			\inferrule[\drule{index-neg}]{
				\dexpr{\Gamma}{\Omega}{e}{r}\\
				r < 0
			}{
				\dexpr{\Gamma}{\Omega}{x[e]}{0}
			}\and
			\inferrule[\drule{cond-undef}]{
				\dexpr{\Gamma}{\Omega}{e_1}{\indef}\\
			}{
				\dexpr{\Gamma}{\Omega}{\eif\;e_1\;\ethen\;e_2\;\eelse\;e_3}{\indef}
			}\and
			\inferrule[\drule{index-undef}]{
				\dexpr{\Gamma}{\Omega}{e}{\indef}\\
			}{
				\dexpr{\Gamma}{\Omega}{x[e]}{\indef}
			}\and
			\inferrule[\drule{index-outside}]{
				\dexpr{\Gamma}{\Omega}{e}{r}\\
				r \geqslant n\\
                                |\Omega(x)| = n
			}{
				\dexpr{\Gamma}{\Omega}{x[e]}{\indef}
			}
			\and
                        \inferrule[\drule{tab-undef}]{
				x \not \in \text{dom }\Omega
			}{
				\dexpr{\Gamma}{\Omega}{x[e]}{\indef}
			}\and
			\inferrule[\drule{index}]{
				\Omega(x) = (v_0, \ldots, v_{n-1})\\
				\dexpr{\Gamma}{\Omega}{e}{r}\\
				r \in [0, n)\\
				r' = \text{truncate}_{\F}(r)
			}{
				\dexpr{\Gamma}{\Omega}{x[e]}{v_{r'}}
			}
			\and
			\inferrule[\drule{func}]{
				\dexpr{\Gamma}{\Omega}{e_1}{v_1}\\
				\cdots\\
				\dexpr{\Gamma}{\Omega}{e_n}{v_n}
			}{
				\dexpr{\Gamma}{\Omega}{f(e_1,\ldots,e_n)}{f(v_1,\ldots,v_n)}
			}
		\end{mathpar}
	}
	\caption{Operational semantics: expressions}
        \label{fig:mum:sem:expr}
\end{figure}
\begin{figure}
	{\footnotesize
		\centering
		\[
			\text{Judgment : }
			\boxed{\dcmd{\Gamma}{\Omega_c}{c}{\Omega'_c}}
			\text{ and }
		\]
		\[
			\boxed{\dprog{\Gamma}{\Omega_c}{P}{\Omega'_c}}
			\quad\text{(\enquote{Under $\Omega_c$, $P$ produces $\Omega'_c$})}
		\]
		\begin{mathpar}
			\inferrule[\drule{assign}]{\Omega_c \neq \lit*{\errorlbl}\\\dexpr{\Gamma}{\Omega_c}{e}{v}}{
				\dcmd{\Gamma}{\Omega_c}{x\;\lit*{:=}\;e}{\Omega_c[x \mapsto v]}
			}
			\and
			\inferrule[\drule{assert-other}]{
				\Omega_c \neq \lit*{\errorlbl}\\
				\dexpr{\Gamma}{\Omega_c}{e}{v}\\v\in\{0,\indef\}
			}{
				\dcmd{\Gamma}{\Omega_c}{\eif\;e\;\ethen\;\eerror}{\Omega_c}
			}\and
			\inferrule[\drule{assert-true}]{
				\Omega_c \neq \lit*{\errorlbl}\\
				\dexpr{\Gamma}{\Omega_c}{e}{f}\\
				f \notin \{0, \indef\}
			}{
				\dcmd{\Gamma}{\Omega_c}{\eif\;e\;\ethen\;\eerror}{\lit*{\errorlbl}}
			}\and
			\inferrule[\drule{error}]{
				\quad
			}{\dprog{\Gamma}{\lit*{\errorlbl}}{c}{\lit*{\errorlbl}}}
			\and
			\inferrule[\drule{seq}]{
				\dcmd{\Gamma}{\Omega_{c,0}}{c}{\Omega_{c,1}}\\
				\dprog{\Gamma}{\Omega_{c,1}}{P}{\Omega_{c,2}}
			}{
				\dprog{\Gamma}{\Omega_{c,0}}{c\; \lit*{;} \; P}{\Omega_{c,2}}
			}
			\and
			\inferrule[\drule{assign-table}]{
				\Omega_c \neq \lit*{\errorlbl}\\
				\dexpr{\Gamma}{\Omega_c[\lit*{X} \mapsto 0]}{e}{v_0}\\
				\cdots\\
				\dexpr{\Gamma}{\Omega_c[\lit*{X} \mapsto n-1]}{e}{v_{n-1}}
			}{
				\dcmd{\Gamma}{\Omega_c}{x[\lit*{X},n]\;\lit*{:=}\;e}{\Omega_c[x \mapsto (v_0,\ldots, v_{n-1})]}
			}
		\end{mathpar}}
	\caption{Operational semantics: statements}
        \label{fig:mum:sem:stmts}
\end{figure}

\begin{figure*}
	{\footnotesize
          \centering
          \begin{minipage}[t]{0.65\textwidth}
            \begin{mathpar}
              \begin{array}{c|cc}
                e_1 \odot e_2, \odot \in \{ +, -\}
                & \indef      & f_2 \in \F    \\
                \hline
                \indef     & \indef      & 0 \odot f_2   \\
                f_1 \in \F & f_1 \odot 0 & f_1 \odot_\F f_2 \\
              \end{array}
              \quad
              \begin{array}{c|ccc}
                e_1 \odot e_2, \odot \in \{\times, \div \} & \indef & f_2 \in \F, f_2 \neq 0 & 0 \\
                \hline
                \indef                                     & \indef & \indef & \indef                  \\
                f_1                                        & \indef & f_1 \odot_\F f_2 & 0\\
              \end{array}\\
              \begin{array}{c|cc}
                b_1 \;\boolop\; b_2 & \indef & f_2 \in \F           \\
                \hline
                \indef               & \indef & \indef               \\
                f_1 \in \F           & \indef & f_1 \;\boolop_\F\; f_2
              \end{array}
              \quad
              \begin{array}{c|cc}
                m(e_1, e_2), m \in \{ \fmin, \fmax \}
                & \indef      & f_2 \in \F    \\
                \hline
                \indef     & 0            & m_\F(0, f_2)  \\
                f_1 \in \F & m_\F(f_1, 0) & m_\F(f_1, f_2) \\
              \end{array}
            \end{mathpar}
          \end{minipage}\qquad
          \begin{minipage}[t]{0.3\textwidth}
            \begin{mathpar}
              \begin{array}{l}
                \farr(\indef) = \indef\\
                \farr(f \in \F) = \text{floor}_{\F}(f + \texttt{sign}(f) * 0.50005)\\
                \finf(\indef) = \indef\\
                \finf(f \in \F) = \text{floor}_{\F}(f + 10^{-6})\\
                \fabs(\lit*{x}) \equiv \eif\; \lit*{x >= 0} \; \ethen \; \lit*{x} \; \eelse \; \lit*{-x}\\
                \fposo\;(\lit*{x}) \equiv \lit*{x >= 0}\\
                \fpos(\lit*{x}) \equiv \lit*{x > 0}\\
                \fnull(\lit*{x}) \equiv \lit*{x = 0}\\
                \fpresent(\indef) = 0\\
                \fpresent(f \in \F) = 1
              \end{array}
          \end{mathpar}
          \end{minipage}
        }
        \caption{Function semantics. {\smaller For context on \texttt{round} and \texttt{truncate} definitions, see \sref{mlang:backends}}}
	\label{fig:mum:sem:func}
\end{figure*}

\myparagraph{Statements}
The memory environment $\Omega$ is extended into $\Omega_c$, to propagate the
error case that may be raised by exceptions. An assignment updates a
valid memory environment with the computed value.  If an assertion's guard
evaluates to a non-zero float, an error is raised;
otherwise, program execution continues.  Rule \TirName{\drule{error}}
propagates a raised error across a program.  The whole-array assignment works
by evaluating the expression in different memory environments, one for each
index.

\subsection{Type safety}

We now prove type safety in Coq. Owing to the unusual
semantics of the \indef value, and to the lax treatment of undefined
variables, this provides an additional level of guarantee, by ensuring that
reduction always produces a value or an error (i.e. we haven't forgotten any
corner cases in our semantics). Furthermore, we show in the
process that the store is consistent with the typing
environment, written $\Gamma \rhd \Omega$. This entails store typing (i.e. values of the right type are to be
found in the store) and proper handling of undefined variables (i.e. $\text{dom
}\Omega \subseteq \text{dom }\Gamma$).

\begin{thm}[Expressions]
  If $\Gamma \rhd \Omega$ and $\texpr{\Gamma}{e}$, then there exists $v$
  such that $\dexpr{}{\Gamma}{e}{v}$.%
\end{thm}

\noindent
We extend $\rhd$ to statements, so as to account for exceptions:
\[\Gamma \rhd_c \Omega_c \Longleftrightarrow \Omega_c =
\lit*{\errorlbl} \vee \Gamma \rhd \Omega_c \]

\begin{thm}[Statements]
  If $\jcmd{\Gamma}{c}{\Gamma'}$ et $\Gamma \rhd_c \Omega_c$, then there exists
  $\Omega_c'$ such that $\dcmd{}{\Omega_c}{c}{\Omega_c'}$ and $\Gamma' \rhd_c
  \Omega_c'$.
\end{thm}

\noindent
We provide full proofs and definitions in Coq, along with a guided tour of our
development, in the supplement \cite{mlang-zenodo}.

\section{The Design of a New DSL: \mpp}
\label{sec:mpp}
\newcommand{\fundeclbl}{fundecl}
\newcommand{\fundec}{\synt{\fundeclbl}}

\newcommand{\commandlbl}{command}
\newcommand{\command}{\synt{\commandlbl}}

\newcommand{\variablekindlbl}{var\_kind}
\newcommand{\kindcallbl}{kind\_filtering}

\newcommand{\funlbl}{funname}
\newcommand{\fcast}{\texttt{cast}}
\newcommand{\funclbl}{fun}
\newcommand{\bltinlbl}{builtin}

\newcommand{\callm}{\texttt{call\_m}}
\newcommand{\kcexists}{\texttt{exists}}
\newcommand{\kcvarhas}{\texttt{var\_has}}

\begin{figure}
  {\footnotesize
	\begin{grammar}
		<\programlbl> ::= <\fundeclbl>*

		<\fundeclbl> ::= <\funlbl> "(" <\varlbl>* "):" <\commandlbl>*

                <\commandlbl> ::= "\iflbl" <\exprlbl> "then" <\commandlbl>* else <\commandlbl>*
                \alt "partition with" <\variablekindlbl> ":" <\commandlbl>*
                \alt <\varlbl> "=" <\exprlbl> | <\varlbl>* "<-" <\funclbl>"()" | "del" <\varlbl>

		<\exprlbl> ::= <\varlbl> | <\floatlbl> | "\indef" | <\exprlbl> <\binoplbl>  <\exprlbl> | <\unoplbl> <\exprlbl>
                \alt  <\bltinlbl> "(" <\exprlbl>, \ldots, <\exprlbl> ")" | "\kcexists(" <\variablekindlbl> ")"

		<\binoplbl> ::= <\arithoplbl>  | <\booloplbl>

		<\arithoplbl> ::= "+" | "-" | "*" | "/"

		<\booloplbl> ::= "<=" | "<" | ">" | ">=" | "==" | "!=" | "&&" | "||"

		<\unoplbl> ::= "-" | "~"

                <\variablekindlbl> ::= "taxbenefit" | "deposit" | ...

                <\funclbl> ::=  <\funlbl> | "\callm"

		<\bltinlbl> ::= "\fpresent" | "\fcast"
              \end{grammar}
            }
	\caption{Syntax of the \mpp language}
        \label{fig:mpp:syntax}
\end{figure}

As described in \fref{dgfip:architecture}, the internal compiler of the
DGFiP compiles M files (\sref{semantics}) to C code. Insofar as we understand,
the M codebase originally expressed
the whole income tax computation. However, in the 1990s (\sref{intro}),
the \dgfip started executing the M code twice, with slightly different
parameters, in order for the taxpayer to witness the impact of a tax reform.
Rather than extending M with support for user-defined functions, the \dgfip
wrote the new logic in C, in a folder called \enquote{inter}, for multi-year
computations.
This piece of code can read and write variables
used in the M codebase using shared global state. To assemble the final
executable, M-produced C files and hand-written \enquote{inter} C files are compiled
by GCC and distributed as a shared library.
Over time, the \enquote{inter} folder grew to handle a variety of special cases,
multiplying calls into the M codebase. At the time of writing, the \enquote{inter}
folder amounts to 35,000 lines of C code.

This poses numerous problems. First, the mere fact that \enquote{inter} is written in
C prevents it from being released to the public, the \dgfip fearing security
issues that might somehow be triggered by malicious inputs provided by the
taxpayer. Therefore, the taxpayer cannot reproduce the tax computation since key
parts of the logic are missing. Second, by virtue of being written in C,
\enquote{inter} does not compose with M, hindering maintainability, readability and
auditability.
Third, C limits the ability to modernize the codebase; right now, the online tax
simulator is entirely written in C using Apache's CGI feature (including HTML
code generation), a very
legacy infrastructure for Web-based development.
Fourth, C is notoriously hard to analyze, preventing both the \dgfip and the
taxpayer from doing fine-grained analyses.%

To address all of these limitations, we design \mpp, a companion domain-specific
language (DSL) that is
powerful enough to completely eliminate the hand-written C code.

\subsection{Concrete syntax and new constructions}

The chief purpose of the \mpp DSL is to repeatedly call the M rules, with
different M variable assignments for each call. To assist with this task, \mpp
provides basic computational facilities, such as functions and local variables.
In essence, \mpp allows implementing a ``driver'' for the M code.

\fref{mpp:example} shows concrete syntax for \mpp. We chose syntax
resembling Python, where block scope is defined by indentation.
As the French administration moves towards a modern digital
infrastructure, Python seems to be reasonably understood across various
administrative services.

\fref{mpp:syntax} formally lists all of the language constructs that \mpp provides. A
program is a sequence of function declarations. \mpp features two flavors of
variables.
Local variables follow scoping rules similar to Python: there is one
local variable scope per function body; however, unlike Python, we disallow
shadowing and have no block scope or \li+nonlocal+ keyword.
Local variables exist only in \mpp.
Variables in all-caps live in the M variable scope, which is shared between M
and \mpp, and obey particular semantics.

\subsection{Semantics of \mpp}

Two constructs support the interaction between M and \mpp:
the \li+<-+ and \li+partition+ operators. They
have slightly unusual semantics, in the way that they deal
with the M variable scope. These semantics are heavily influenced by the
needs of the \dgfip, as we strived to provide something that would feel
intuitive to technicians in the French administration.

To precisely define the expected behavior,
\fref{mpp:reduction} presents reduction semantics of the form $\Delta, \Omega_1 \vdash c
\rightsquigarrow \Omega_2$, meaning command $c$ updates
the store from $\Omega_1$ to $\Omega_2$, given the functions declared in $\Delta$.

We distinguish built-ins, which may only appear in expressions
and do not modify the global store, from functions, which are declared at the
top-level and may modify the store. The \li+call_m+ operation is a special
function. The \li+<-+ operator takes a
\emph{function} call, and executes it in a copy of the memory. Then, only those
variables that appear on the left-hand side see their value propagated to the
parent execution environment. Thus, \li+call_m+ only affects variables $\vec X$.

To execute the function call, the \li+<-+ operator either looks up definitions
in $\Delta$, the environment of user-defined functions, or executes the M rules
in the \li+call_m+ case, relying on the earlier definition of $\Rrightarrow$
(\fref{mum:sem:stmts}).

Worded differently, our semantics introduce a notion of call stack and treat the
M computation as a function call returning multiple values. It is to be noted
that the original C code had no such notion, and that the $\vec X$ were nothing
more than mere comments. As such, there was no way to statically rule out potential hidden
state persisting from one \li+call_m+ to another since the global scope was
modified in place.
With this formalization and its companion
implementation (\sref{exec}), we were able to confirm that there is currently no
reliance on hidden state (something which we suspect took considerable effort to
enforce in the hand-written C code), and were able to design a much more
principled semantics that we believe will lower the risk of future errors.

The \li+partition+ operation operates over a variable kind $k$ (\sref{overview}).
The sub-block $c$ of \li+partition+ executes in a restricted scope, where
variables having kind $k$ are temporarily set to \li+undef+. Upon completion of
$c$, the variables at kind $k$ are restored to their original value, while other
variables are propagated from the sub-computation into the parent scope. This
allows running computations while ``disabling'' groups of variables, e.g.
ignoring an entire category of tax credits.

\begin{figure}
  \begin{Verbatim}[numbers=left,xleftmargin=15pt,fontsize=\footnotesize]
compute_benefits():
 if exists(taxbenefit) or exists(deposit):
    V_INDTEO = 1
    V_CALCUL_NAPS = 1
    partition with taxbenefit:
      NAPSANSPENA, IAD11, INE, IRE, PREM8_11 <- call_m()
    iad11 = cast(IAD11)
    ire = cast(IRE)
    ine = cast(INE)
    prem = cast(PREM8_11)
    V_CALCUL_NAPS = 0
    V_IAD11TEO = iad11
    V_IRETEO = ire
    V_INETEO = ine
    PREM8_11 = prem
\end{Verbatim}
\caption{Example function defined in \mpp}
\label{fig:mpp:example}
\end{figure}

\begin{figure*}[h]
  {\footnotesize
    \[
    \text{Judgments: }
    \boxed{\Delta, \Omega \vdash e \downlsquigarrow v}\quad\text{(\enquote{Under $\Delta, \Omega$, $e$ evaluates into v})}
    \qquad
    \boxed{\Delta, \Omega_1 \vdash c \rightsquigarrow \Omega_2}\quad\text{(\enquote{Under $\Delta$, $c$ transforms $\Omega_1$ into $\Omega_2$})}
  \]
  \centering
  \begin{mathpar}
    \inferrule[\textsc{Cast-float}]
    {
      \Delta, \Omega \vdash \textsf{e} \downlsquigarrow f\\
      f \neq \indef
    }
    {
      \Delta, \Omega \vdash \textsf{cast(e)} \downlsquigarrow f
    }
    \and
    \inferrule[\textsc{Cast-undef}]
    {
      \Delta, \Omega \vdash \textsf{e} \downlsquigarrow \indef
    }
    {
      \Delta, \Omega \vdash \textsf{cast(e)} \downlsquigarrow 0
    }
    \and
    \inferrule[\textsc{Exists-true}]
    {
      \exists X \in \Omega, kind(X) = k \wedge \Omega(X) \neq \indef
    }
    {
      \Delta, \Omega \vdash \textsf{exists(k)} \downlsquigarrow 1
    }
    \and
    \inferrule[\textsc{Exists-false}]
    {
      \forall X \in \Omega, kind(X) \neq k \vee \Omega(X) = \indef
    }
    {
      \Delta, \Omega \vdash \textsf{exists(k)} \downlsquigarrow 0
    }
    \and
    \inferrule[
      \textsc{Call}
    ]{
      \begin{array}{ll}
        \phantom{\Delta,\,}\Omega_1 \vdash \textsf{M rules} \Rrightarrow \Omega_2 & \text{if }
        \textsf{f} = \textsf{call\_m } \cr
        \Delta, \Omega_1 \vdash \Delta(f) \rightsquigarrow \Omega_2 & \text{otherwise }
      \end{array} \\
      \begin{array}{l}
        \Omega_3(Y) = \Omega_1(Y) \text{ if } Y \not \in \vec X \cr
        \Omega_3(Y) = \Omega_2(Y) \text{ if } Y \in \vec X
      \end{array}
    }{
    \Delta, \Omega_1 \vdash \vec X \leftarrow \textsf{f}() \rightsquigarrow \Omega_3
    }
    \and
    \inferrule[
      \textsc{Partition}
    ]{
      \begin{array}{l}
        \Omega_2(Y) = \textsf{undef} \text{ if } \textsf{kind}(Y) = k \cr
        \Omega_2(Y) = \Omega_1(Y)\text{ otherwise }
      \end{array} \\
      \Delta, \Omega_2 \vdash c \rightsquigarrow \Omega_3 \\
      \begin{array}{l}
        \Omega_4(Y) = \Omega_1(Y) \text{ if } \textsf{kind}(Y) = k \cr
        \Omega_4(Y) = \Omega_3(Y) \text{ otherwise }
      \end{array} \\
    }{
    \Delta, \Omega_1 \vdash \textsf{partition with k}: c \rightsquigarrow \Omega_4
    }
    \and
    \inferrule[
      \textsc{Delete}
      ]
      {
        \Delta, \Omega_1 \vdash \textsf{v = undef} \rightsquigarrow \Omega_2
      }
      {
        \Delta, \Omega_1 \vdash \textsf{del v} \rightsquigarrow \Omega_2
      }
  \end{mathpar}
  \caption{Reduction rules of \mpp}
  \label{fig:mpp:reduction}
  }
\end{figure*}

\subsection{Example}
\label{sec:mpp:example}

\fref{mpp:example} provides a complete \mpp example, namely the function
\li+compute_benefits+.

The conditional at line 2 uses a variable kind-check (\sref{overview}) to see
if any variables of kind \enquote{tax benefit} have a non-\li+undef+ value.
Then, lines 3-4 set some flags before calling M. Line 5 tells us that
the call to M at line 6 is to be executed in a restricted context where variables of
kind \enquote{tax benefit} are set to \li+undef+.
Line 6 runs the M computation, over the current state of the M variables; five M output
variables are retained from this M execution, while the rest are discarded.
Lines 7-11 represent local variable assignment, where \li+cast+ has the same
effect as \li!+ 0! in M, namely, forcing the conversion of \li+undef+ to 0.
Then, lines 11-15 set M some variables as input for later function calls.

\section{\mlang: an M/\mpp Compiler}
\label{sec:exec}

After clarifying the semantics of M (\sref{semantics}), and designing a new DSL
to address its shortcomings (\mpp, \sref{mpp}), we now present
\mlang, a modern compiler for both M and \mpp.

\subsection{Architecture of \mlang}

\mlang takes as input an M codebase, an \mpp file, and a file specifying
assumptions (described in the next paragraph).  \mlang currently generates
Python or C; it also offers a built-in interpreter for computations.
\mlang is implemented in \ocaml, with
around 9,000 lines of code.  The general architecture
is shown in \fref{mlang:structure}.  The M files and the \mpp program
are first parsed and transformed into intermediate representations.  These
intermediate representations are inlined into
a single backend intermediate representation (BIR), consisting of assignments
and conditionals. Inlining is aware of the semantic subtleties described in
\fref{mpp:reduction} and uses temporary variable assignments to save/restore the
shared M/\mpp scope.
BIR code is then translated to the optimization intermediate
representation (OIR) in order to perform optimizations. OIR is the
control-flow-graph (CFG) equivalent of BIR.

OIR is the representation on which we perform our optimizations
(\sref{mlang:opt}).
For instance, in order to perform constant propagation, we must check that a given assignment
to a variable dominates all its subsequent uses. A CFG is the best data
structure for this kind of analysis. We later on switch back to the AST-based
BIR in order to generate textual C output.

\begin{figure*}%
	\begin{center}
          \scalebox{1}{
            \begin{tikzpicture}[scale=1,font=\smaller]
              \node (msource) [Box] {\texttt{sources.m}};
              \node (mppsource) at ($ (msource)+(0,-1) $) [Box] {\texttt{source.mpp}};
              \node (mast)  at ($ (msource)+(3,0)$) [Box] {M AST};
              \node (mppast)  at ($ (mppsource)+(3,0)$) [Box] {\mpp AST};
              \node (mir) at ($ (mast)+(3,0) $) [Box] {M IR};
              \node (mppir) at ($ (mppast)+(3,0) $) [Box] {\mpp IR};
              \node (bir) at ($ (mir)+(3,0) $) [Box] {BIR};
              \node (mspec) at ($ (bir)+(0,1) $) [Box] {\texttt{assumptions.m\_spec}};
              \node (oir) at ($ (bir)+(0,-1) $) [Box] {OIR};
              \node (C) at ($ (bir)+(2.6,-1) $) [Box] {C};
              \node (Python) at ($ (bir)+(3,1) $) [Box] {Python};
              \node (Interpreter) at ($ (bir)+(3.25,0) $) [Box] {Interpreter};
              \draw[-stealth,color=gray,line width=1pt] (msource) -- (mast);
              \draw[-stealth,color=gray,line width=1pt] (mppsource) -- (mppast);
              \draw[-stealth,color=gray,line width=1pt] (mast) -- (mir);
              \draw[-stealth,color=gray,line width=1pt] (mppast) -- (mppir);
              \draw[-stealth,color=gray,line width=1pt] (mspec) -- (bir);
              \draw[-stealth,color=gray,line width=1pt] (mir) -- (bir);
              \draw[-stealth,color=gray,line width=1pt] (mppir) -- (bir);
              \draw[stealth-stealth,color=gray,line width=1pt] (bir) -- (oir);
              \draw[stealth-stealth,color=gray,line width=1pt] ($ (oir)+(-0.225,-0.25) $) arc (140:400:0.3);
              \draw[-stealth,color=gray,line width=1pt] (bir) -- (C);
              \draw[-stealth,color=gray,line width=1pt] (bir) -- (Python);
              \draw[-stealth,color=gray,line width=1pt] (bir) -- (Interpreter);
              \node (parsing) at ($ (mppsource)+(1.5,-1) $) { Parsing };
              \draw[loosely dotted, color=red, line width=1] (parsing) -- ($ (parsing)+(0,2.5) $);
              \node (desugaring) at ($ (mppast)+(1.5,-1) $) { Desugaring };
              \draw[loosely dotted, color=red, line width=1] (desugaring) -- ($ (desugaring)+(0,2.5) $);
              \node (inlining) at ($ (mppir)+(1,-1) $) { Inlining };
              \draw[loosely dotted, color=red, line width=1] (inlining) -- ($ (inlining)+(0,2.5) $);
              \node (optimizing) at ($ (oir)+(0,-1) $) { Optimization };
              \node (transpiling) at ($ (bir)+(2,-2) $) { Transpiling };
              \draw[loosely dotted, color=red, line width=1] (transpiling) -- ($ (transpiling)+(0,3) $);
            \end{tikzpicture}}
	\end{center}
	\caption{\mlang compilation passes}
        \label{fig:mlang:structure}
\end{figure*}

\myparagraph{Additional assumptions}
In M, a variable not defined in the current memory environment evaluates to
\indef (rule \TirName{D-Var-Undef}, \fref{mum:sem:expr}).
This permissive behavior is fine for an interpreter which has a dynamic
execution environment; however, our goal is to generate efficient C and Python
code that can be integrated into existing software. As such, declaring every
single one of the 27,113
possible variables (as found in the original M rules) in C would be quite unsavory.

We therefore devise a mechanism that allows stating ahead of time which
variables can be truly treated as inputs, and which are the outputs that we are
interested in. Since these vary depending on the use-case, we choose to list
these assumptions in a separate file that can be provided alongside with the
M/\mpp source code, rather than making this an intrinsic, immutable property set
at variable-declaration time. Doing so increases the quality of the generated C or Python.

We call these \emph{assumption files}; we have hand-written 5 of those.
  \textbf{All} is the empty file, i.e. no additional assumptions. This leaves 2459 input
    variables, and 10,411 output variables for the 2018 codebase.
  \textbf{Selected outs} enables all input variables, but retains only 11 output variables.
  \textbf{Tests} corresponds to the inputs and outputs used in the test files
    used by the DGFiP.
  \textbf{Simplified} corresponds to the simplified simulator released each year
    by the DGFiP a few months before the full income tax computation is
    released. There are 214 inputs, and we chose 11 output variables.
  \textbf{Basic} accepts as inputs only the marital status and the salaries of each
    individual of the couple. The output is the income tax.

\subsection{Optimizations}
\label{sec:mlang:opt}

In the 2018 tax code, the initial number of BIR instructions after inlining M
and \mpp files together is 656,020. This essentially corresponds to what the
legacy compiler normally generates, since it performs no optimizations.

Thanks to its modern compiler architecture, \mlang can easily perform numerous
textbook optimizations, namely dead code elimination, inlining and partial
evaluation. This allows greatly improving the quality of the generated
code.

We now present a series of optimizations, performed on the OIR intermediate
representation.
The number of instructions after these optimizations is shown in \fref{mlang:optimizations}.
Without any assumption (\textbf{All}), the optimizations shrink the generated C code to 15\%
size (a factor of $6.5$). With the most restrictive assumption file
(\textbf{Simplified}), only 0.47\%
optimization.

\begin{figure}
  \begin{tabular}{lrrr}
    \toprule
    Spec. name    & \# inputs & \# outputs & \# instructions\\
    \midrule
    All           & 2,459     & 10,411     & 129,683\\
    Selected outs & 2,459     & 11         &  99,922\\
    Tests         & 1,635     & 646        & 111,839\\
    Simplified    & 228       & 11         &   4,172\\
    Basic         & 3         & 1          &     553\\
    \bottomrule
  \end{tabular}
  \caption{Number of instructions generated after optimization. Instructions
  with optimizations disabled: 656,020.}
  \label{fig:mlang:optimizations}
\end{figure}

\newcommand{\A}[1]{#1^\#}
\newcommand\absorb{$\A{\text{\texttt{absorb}}}$}
\newcommand\cast{$\A{\text{\texttt{cast}}}$}
\newcommand\aindef{\A\indef}
\newcommand\afloat{\A\float}

\myparagraph{Definedness analysis}
Due to the presence of \li+undef+, some usual optimizations are not available.
For example, optimizing \li+e * 0+ into \li+0+ is incorrect when \li+e+ is \li+undef+, as \li+undef * 0 =+
\li+undef+.
Similarly, \li|e + 0| cannot be rewritten as \li+e+.
Our partial evaluation is thus combined with a simple definedness analysis.
The lattice of the analysis is shown in \fref{mlang:definedness:lattice};
we use the standard sharp symbol of abstract interpretation \cite{CC77} to denote abstract elements.
The transfer function \absorb\; defined in \fref{mlang:definedness:funcs} is used to compute the definedness in the case of the multiplication, the division and all operators in $\boolop$.
The \cast\; transfer function is used for the addition and the subtraction.

\begin{figure}
  \begin{tikzpicture}[font=\smaller]
      \node (top) at (0,1) {$\top$};
      \node (indef) at (-1,0) {$\aindef$};
      \node (float) at (1,0) {$\afloat$};
      \node (bot) at (0,-1) {$\bot$};
      \draw (bot) -- (indef) -- (top) -- (float) -- (bot);
    \end{tikzpicture}
    \caption{Definedness lattice}
    \label{fig:mlang:definedness:lattice}
\end{figure}
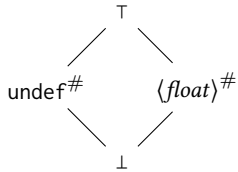
\begin{figure}
  \smaller
    \begin{tabular}{cccc}
      \toprule
      $d_1$ & $d_2$ & \absorb$(d_1, d_2)$ & \cast$(d_1, d_2)$\\
      \midrule
      $\aindef$ & $\aindef$ & $\aindef$ & $\aindef$\\
      $\aindef$ & $\afloat$ & $\aindef$ & $\afloat$\\
      $\afloat$ & $\aindef$ & $\aindef$ & $\afloat$\\
      $\afloat$ & $\afloat$ & $\afloat$ & $\afloat$\\
      \bottomrule
    \end{tabular}
    \caption{Transfer functions over the definedness lattice, implicitly lifted to the full lattice.}
    \label{fig:mlang:definedness:funcs}
\end{figure}

This definedness analysis enables finer-grained partial evaluation rules, such
as those presented in \fref{mlang:optexamples}.

\begin{figure}
  \smaller
  \begin{mathpar}
    \begin{array}{ll}
    e + \indef      \leadsto e&
    e : \afloat + 0 \leadsto e\\
    e * 1           \leadsto e&
    e : \afloat * 0 \leadsto 0\\
    \fmax(0, \fmin(0, x)) \leadsto 0&
    \fpresent(\indef)    \leadsto 0\\
    \fmax(0, -\fmax(0, x)) \leadsto 0&
    \fpresent(e : \afloat)  \leadsto 1\\
    \end{array}
  \end{mathpar}
  \caption{Examples of optimizations}
  \label{fig:mlang:optexamples}
\end{figure}

The optimizations for \li!+ 0! and \li+* 0+ are
invalid in the presence of IEEE-754 special values (NaN, minus zero, infinities)
\cite{fpoptimizations,handbookfp}. We have instrumented the M code to
confirm that these are valid on the values used.
But for safety, these unsafe optimizations are only enabled
if the \li+--fast_math+ flag is set.

\subsection{Backends}
\label{sec:mlang:backends}

\myparagraph{\dgfip (legacy)}
The \dgfip's legacy system has a single backend that produces pre-ANSI (K\&R) C.
For each M rule, two C computations are emitted. The first one aims to determine
whether the resulting value is defined. It operates on C's \li+char+ type, where
\li+0+ is undefined or \li+1+ is defined. The second computation is
syntactically identical, except it operates on \li+double+ and thus computes the
actual arithmetic expression. This two-step process explains some of the
operational semantics: with \li+0+ being undefined, the special value
\li+undef+ is absorbing for e.g. the multiplication.

Careful study of the generated code also allowed us to nail down some
non-standard rounding and truncation rules which had until then eluded us. We
list them in \fref{mlang:customrounding}; these are used to implement the built-in operators from
\fref{mum:syntax} in both our interpreter and backends.

\begin{figure}
{\footnotesize
\begin{verbatim}
// my_var1 is a local variable always defined
#define my_truncate(a)	( my_var1=(a)+0.000001,floor(my_var1) )
#define my_round(a)	(floor(
  (a<0) ? (double)(long long)(a-.50005)
        : (double)(long long)(a+.50005)))
\end{verbatim}
}
\caption{Custom rounding and truncation rules}
\label{fig:mlang:customrounding}
\end{figure}

\myparagraph{\mlang}
Our backend generates C and Python from BIR. Since BIR only features
assignments, arithmetic and conditionals, we plan to extend it with backends for JavaScript, R/MatLab
and even SQL for in-database native tax computation. Depending on the \dgfip's
appetite for formal verification, we may verify the whole compiler since the
semantics are relatively small.

Implementing a new backend is not very onerous: it took us 500 lines for the C
backend and 375 lines for the Python backend. Both backends are validated by
running them over the entire test suite and comparing the result with our
reference interpreter.

Our generated code only relies on a small library of helpers which implement
operations over M values. These helpers are aware of all the semantic subtleties
of M and are manually audited against the paper semantics.

\section{Analyzing and Evaluating the Tax Code}
\label{sec:eval}
Due to the sheer size of the code and number of variables, generating efficient
code is somewhat delicate -- we had the pleasure of breaking both
the Clang and Python parsers because of an exceedingly naïve translation.
Thankfully, owing to our flexible architecture for \mlang, we were able to
quickly iterate and evaluate several design choices.

We now show the benefits of a modern compiler infrastructure, and proceed to
describe a variety of instrumentations, techniques and tweaking knobs that
allowed us to gain insights on the the tax computation. By bringing the M
language into the 21\textsuperscript{st} century, we not only greatly enhance the quality of the
generated code, but also unlock a host of techniques that significantly increase
our confidence in the French tax computation.

\subsection{Performance of the generated code}

We initially generated C code that would emit one local variable per M variable.
But with tens of thousands of local variables,  running the code required \li+ulimit -s+.

We analyzed the legacy code and found out that the \dgfip stored all of the M
variables in a global array. We implemented the same technique and found out
that with \li+-O1+, we were almost as fast as the legacy code. We attribute this
improvement to the fact that the array, which is a few dozen kB, which fits in
the L2 cache of most modern processors. This is a surprisingly fortuitous choice
by the \dgfip. 
See \fref{tooling:perfresults} for full results. 
In the grand scheme of things, the cost of computing the final 
tax is dwarfed by the time spent generating a PDF summary for the taxpayer 
($\sim$200ms). The 500µs difference between the \dgfip's system and ours is 
thus insignificant.

\begin{figure}
\small
\begin{tabular}{lcccc}
\toprule
Scheme&M compiler&C compiler&Bin. size&Time\\\midrule
Original&DGFiP&GCC \texttt{-O0}&7 Mo&$\sim 1.5$ ms\\
Original&DGFiP&GCC \texttt{-O1}&7 Mo&$\sim 1.5$ ms\\
Array&\mlang&Clang \texttt{-O0}&19 Mo&$\sim 4$ ms\\
Array&\mlang&Clang \texttt{-O1}&10 Mo&$\sim 2$ ms\\
\bottomrule
\end{tabular}
\caption{
    Performance of the C code generated by various compilation schemes for the M code.
    {\smaller
    The time measured is the time spent inside the main tax computation function for
    one fiscal household picked in the set of test cases. Size of the compiled
    binary is indicated. \enquote{Original}
    corresponds to the DGFiP's legacy system. \enquote{Local vars} corresponds
    to \mlang's C backend mapping each M variable to a C local variable.}
}
\label{fig:tooling:perfresults}
\end{figure}

\subsection{The cost of IEEE-754}
\label{sec:ieee754}

Relying on IEEE-754 and its limited precision for something as crucial as the
income tax of an entire nation naturally raises questions. Thanks to our new
infrastructure, we were able to instrument the generated code and gain numerous
insights.

\mypointlessparagraph{Does precision matter?}
We tweaked our backend to use the MPFR multiprecision library~\cite{fousse2007mpfr}. With
1024-bit floats, all tests still pass, meaning that there is no loss of
precision with the double-precision 64-bit format.

\mypointlessparagraph{Does rounding matter?}
We then instrumented the code to measure the effect of the IEEE-754 rounding
mode on the final result. Anything other than the default (rounding to nearest,
ties to even) generates incorrect results. The control-flow remains roughly the
same, but some comparisons against 0 do give out different results as the computation
skews negatively or positively. We plan in the future to devise a static
analysis that could formally detect errors, such as comparisons that are always false,
or numbers that may be suspiciously close to zero (denormals).

\myparagraph{Fixed precision}
Nevertheless, floating-point computations are notoriously hard to analyze and
reason about, so we set out to investigate replacing floats with integer values.
In our first experiment, we adopted big decimals, i.e. a bignum for the integer
part and a fixed amount of digits for the fractional part. Our test suite
indicates that the integer part never exceeds 9999999999 (encodable in 37 bits);
it also indicates that with 40 bits of precision for the fractional part, we get
correct results. This means that a 128-bit integer would be a viable alternative
to a \li+double+, with the added advantage that formal analysis tools would be
able to deal with it much better.

\myparagraph{Using rationals}
Finally, we wondered if it was possible to completely work without
floating-point and eliminate imprecision altogether, taking low-level details
such as rounding mode and signed zeroes completely out of the picture.

To that end, we encoded values as fractions where both numerator and denominator
are big integers. We observed that both never exceed $2^{128}$, meaning we could
conceivably implement values as a struct with two 128-bit integers and a sign
bit. We have yet to investigate the performance impact of this change.

\subsection{Test-case generation}
\label{sec:testcases}

The \dgfip test suite is painstakingly constructed by hand by lawyers year after year.
From this test suite, we extracted 476 usable test cases that don't
raise any exceptions (see \sref{overview}).
The \dgfip has no infrastructure to automatically generate cases that would
exercise new situations.
As such, the test suite remains relatively limited in the variety of households
it covers. Furthermore, many of the hand-written tests are for previous editions
of the tax code, and describe situations that would be rejected by the current
tax code.

Generating test cases is actually non-trivial: the search space is incredibly
large, owing to the amount of variables, but also deeply constrained, owing to
the fact that most variables only admit a few possible values (\sref{intro}),
and are further constrained in relationship to other variables.

We now set out to automatically generate fresh (valid) test cases for the tax
computation, with two objectives: assert on a very large number of test cases
that our code and the legacy implementation compute the same result; and exhibit
corner cases that were previously not exercised, so as to generate fresh
novel tax situations for lawmakers to consider.

\myparagraph{Randomized testing}
We start by randomly mutating the legacy test suite, in order to generate
new distinct, valid test cases. If a test case raises an exception, we discard
it. We obtain 1267 tests, but these are, unsurprisingly, very close to the
legacy test suite and do not exercise very many new situations. They did,
however, help us when reverse-engineering the semantics of M. We now have 100\%
conformance on those tests.

\myparagraph{Coverage-guided fuzzing} In order to better explore the search space,
we turn to AFL~\cite{afl}. The tool admits several usage modes -- finding
genuine crashes (e.g. segfaults), or generating test cases for further seeding
into the rest of the testing pipeline.  We focus on the latter mode, meaning
that we generate an artificial \enquote{crash} when a synthesized testcase raises no M
errors, that is, when we have found a valid testcase.
We first devise an injection from opaque binary inputs, which AFL controls,
to the \dgfip input variables. Once \enquote{crashes} have been collected, we simply
emit a set of test inputs that has the same format as the \dgfip.

Thanks to this very flexible architecture, we were able to perform fully general
fuzzing exercising all input variables, as well as targeted fuzzing that focuses
on a subset of the variables.  The former takes a few hours on a
high-end machine; the latter mere minutes. We synthesized around 30,000 tests
cases, which we reduced down to 275 using afl-cmin.

So far, the fuzzer-generated test case have pointed out of a few bugs in
\mlang's optimizations and backends. We plan to further use AFL to find
find test cases that satisfy extra properties
not originally present in the tax code, e.g. an excessively high marginal tax
rate that might raise some legality questions.

\myparagraph{Symbolic execution fuzzing}
We attempted to use dynamic symbolic execution tool KLEE~\cite{klee}, but found
out that it only had extremely limited support for floating-point computations.
As detailed earlier (\sref{ieee754}), we have found that integer based
computations are a valid replacement for floats, and plan to use this alternate
compilation scheme to investigate whether KLEE would provide interesting test
cases.

\subsection{Coverage measurements}

Finally, we wish to evaluate how \enquote{good} our new test cases are. Code coverage seems
like a natural notion, especially seeing that there is currently none in the \dgfip
infrastructure. However, traditional code coverage makes little sense:
conditionals are very rare in the generated code.

Rather, we focus on value coverage:
for each assignment in the code, we count the number of distinct values
assigned during the execution of an entire test case. This is a good proxy
for test quality: the more different values flow through an assignment, the more
interesting the tax situation is.

\fref{coverage} shows our measurements. The first take-away is that our
randomized tests did not result in meaningful tests: the number of assignments
that are uncovered actually increased. The tests we obtained with AFL, however,
significantly increase the quality of test coverage. We managed to synthesize
many tests that exercise statements previously unvisited by the \dgfip's test
suite, and exhibit much more complex assignments (2 or more different values
assigned).

Our knowledge of the existing \dgfip test suite is incomplete, as we only have
access to a partial set of tests. In particular, a special set of rules apply
when the tax needs to be adjusted later on following an audit, and the tests for
these have not been communicated to us. We hope to obtain visibility onto those
in the future.

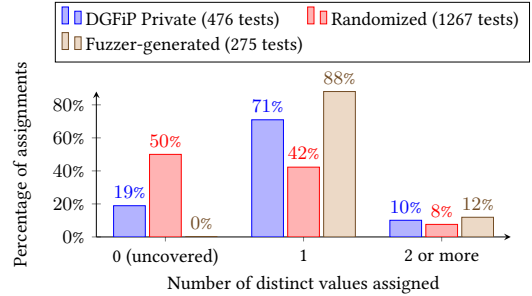
\begin{figure}
\begin{tikzpicture}[scale=.8,font=\smaller]
\begin{axis}[
    ybar,
    xlabel=Number of distinct values assigned,
    ylabel=Percentage of assignments,
    width=\linewidth,%
    height=4cm,
    bar width=15pt,
    xmin=-0.5,
    xmax=2.5,
    ymin=0,
    axis lines=left,
    legend cell align=left,
    legend style={at={(-0.1,1.2)},anchor=south west},
    nodes near coords={\pgfmathprintnumber[precision=0,fixed]{\pgfplotspointmeta}\%},
    legend columns=2,
    yticklabel={\pgfmathprintnumber{\tick}\%},
    xtick={0,1,2},
    xticklabels={0 (uncovered),1,2 or more},
]
\addplot
    table[x=Number of values,y=Percentage of assignments,col sep=comma]
    {coverage2018privatetests.table};
\addplot
    table[x=Number of values,y=Percentage of assignments,col sep=comma]
    {coverage2018publicrandomizedtests.table};
\addplot
    table[x=Number of values,y=Percentage of assignments,col sep=comma]
    {coverage2018fuzzertests.table};
\legend{DGFiP Private (476 tests),Randomized (1267 tests), Fuzzer-generated (275 tests)}
\end{axis}
\end{tikzpicture}
\caption{Value coverage of assignments for each test suite}
\label{fig:coverage}
\end{figure}

\section{Related Work and Conclusion}
\label{sec:conclusion}
\subsection{Implementing the law}

Formalizing part of the law using logic programming or a custom domain
specific language has been extensively tried in the past, as early as 1914
\cite{dewey1914logical, allen1956symbolic,
normalizedlegaldrafting,nationalityact, drivingformalized, holzenberger2020dataset, pertierra2017towards}.
Most of these works follow the same structure: they take a subset of the law,
analyze its logical structure, and encode it using a novel or existing
formalism. All of them stress the complexity of this formal endeavor, coming
from \emph{i)} the underlying reality that the law models and \emph{ii)} the
logical structure of the legislative text itself. After more a century of research,
no silver bullet has emerged that would allow to systematically translate the
text of a law into a formal model.

However, domain-specific attempts have been more successful. Recently, blockchain
has demonstrated increased interest for domain-specific languages encoding smart contracts
\cite{hvitved2011contract, scoca2017smart, he2018spesc, zakrzewski2018towards}.
Regular private commercial contracts have also been targeted for formalization
\cite{l4legalese, camilleri2017contracts}, as well as financial contracts \cite{
eber2009financial, imandra}. Concerning the public sector,
the \enquote{rules as code} movement has been the object of an exhaustive OECD
report \cite{racprimer}.

Closer to the topic of this paper, the logical structure of the US tax law has
been extensively studied by Lawsky \cite{lawsky2017, lawsky2018}, pointing out
the legal ambiguities in the text of the law that need to be resolved using
legal reasoning. She also claims that the tax law drafting style follows default
logic \cite{Reiter1987}, a non-monotonic logic that is hard to encode in languages
with first-order logic (FOL). This could explain, as M is also based on FOL,
the complexity of the \dgfip codebase.

As this complexity generates opacity around the way taxes are computed, another
government agency set out to re-implement the entire French socio-fiscal
system in Python \cite{shulz2019free}. Even if this initiative was helpful and
used as a computation backend for various online simulators, the results it
returns are not legally binding, unlike the results returned by the \dgfip.
Furthermore, this Python implementation does not deal with all the corner cases
of the law.
To the extent of our knowledge, our work is unprecedented in terms of size and
exhaustiveness of the portion of the law turned into a reusable and formalized
software artifact.

\subsection{Conclusion}

Thanks to modern compiler construction techniques, we have been able to lift up
a legacy, secret codebase into a reusable, public artifact that can be distributed
into virtually any programming environment. The natural next step for the \dgfip is
to consider taking more insight from programming languages research, and design
a successor to M/M++ that provides good tooling for translating the tax law
into a correct and distributable implementation.

\begin{acks}
  This work is partially supported by the
  \grantsponsor{ERC}{European Research Council}{https://erc.europa.eu/} under 
  Consolidator Grant Agreement \grantnum{ERC}{681393} --- MOPSA and \grantnum{ERC}{683032} --- CIRCUS.
\end{acks}
\bibliography{../../../verifisc}

\end{document}